# A Questionable Fundamental Basis for Quantum Computing Advantage

Stuart Mirell[*] and Daniel Mirell

*Quantum Wave Technologies, Los Angeles, CA, USA; https://quwt.com*

[*]*email: smirell@ucla.edu*

(Dated: June 4, 2022)

**ABSTRACT**

The widely accepted basis for quantum computing advantage is derived from the entanglement and superposition properties of the probabilistic interpretation of the underlying quantum mechanical formalism which in turn is widely accepted based upon results of Bell experiments. That advantage is questioned here in view of a locally real interpretation that is not negated by Bell experiments and under which entanglement and superposition are excluded.

**INTRODUCTION**

We consider here the viability of quantum advantage in which a quantum computer utilizing entanglement and superposition exhibits speedup relative to a conventional computer by exploiting the uniquely "quantum mechanical" properties of entanglement and superposition postulated by the widely accepted probabilistic interpretation, PI. Because of that widespread acceptance of PI and the general recognition that calculational methods such as Shor's algorithm [1] are valid given PI, quantum advantage is regarded as valid in principle. Difficulties to date in definitively demonstrating quantum advantage are then commonly attributed to technical difficulties in coping with decoherence of the quantum entities used to perform the calculations.

In this report we propose that the difficulties in demonstrating quantum advantage are far more fundamental and those difficulties relate instead to the viability of PI itself. This is certainly a controversial proposition given that the results of Bell's theorem [2] and performed Bell experiments such as [3,4] are virtually universally regarded as excluding alternatives to PI such as locally real hidden variable theories, HVT's.

In this regard we examine a particular locally real representation, denoted as "LR", that is identifiable as an HVT. In this examination our focus is on the viability of LR by clearly demonstrating the absence of entanglement (non-local quantum correlation) and superposition in a particular locally real representation that is differentially testable against PI.

In ref. [5], a detailed extensive basis is presented for the LR representation with emphasis on the exclusion of entanglement. The rationale for the validity of that basis is briefly reviewed here. The ref. [5] derivation of that basis from first principles is not repeated in this report. However, a simplified model of that basis given in [5] is included here in the interests of succinctly demonstrating the locality of quantum correlation for the LR representation. The ref. [5] formulation of LR primarily treats "observables", meaning quantum waves occupied by energy quanta as opposed to empty waves on which no energy quanta reside but which are nonetheless objectively real in LR.

In the present report we extend the ref. [5] LR treatment of observables with a new, detailed analysis of photon wave packets, both occupied and empty, and their interactions with physical systems such as polarizers, calcite loops and Mach-Zehnder polarizing beam splitter loops. In this analysis we show that the property of superposition, is explicitly absent in the LR representation.

**ENTANGLEMENT**

Perhaps the greatest impediment to the formulation of a viable locally real representation of quantum mechanics is the perceived exclusion of "locally real hidden variable theories" (HVT's) based upon results of performed experiments in conjunction with Bell's Theorem. [2]

In that regard, the particular LR representation of local realism ref. [5] derived from first principles of the underlying quantum formalism is shown to be in exact agreement with PI and in agreement with results of performed "Bell" experiments. The LR representation is inherently inclusive of a 3-dimentional wave structure for photons and a 3-dimentional wave structure for particles. Based on the parameters of those wave structures, LR is appropriately categorized as a locally real hidden variable theory, HVT.

For example, in LR the objectively real longitudinal aspect of photon wave structure (along the propagation axis) is represented by the wave function of the underlying quantum mechanical formalism. However, critically with respect to hidden variables, that longitudinal aspect is shown to span an arc in the transverse plane. For any given photon, the magnitude of the arc span and the orientation of the arc bisector are appropriately regarded as the hidden variables attributable to that photon. For virtually all commonly generated photons, the magnitude of that arc span is a "full complement" $\Delta=\pi/2$. The arc bisector orientation $\theta$ fully determines the measurement outcome of that photon traversing a polarizer for which the polarization axis is specified. In the example depicted in Fig. 1, the arc bisector of the objectively real photon is -40°.

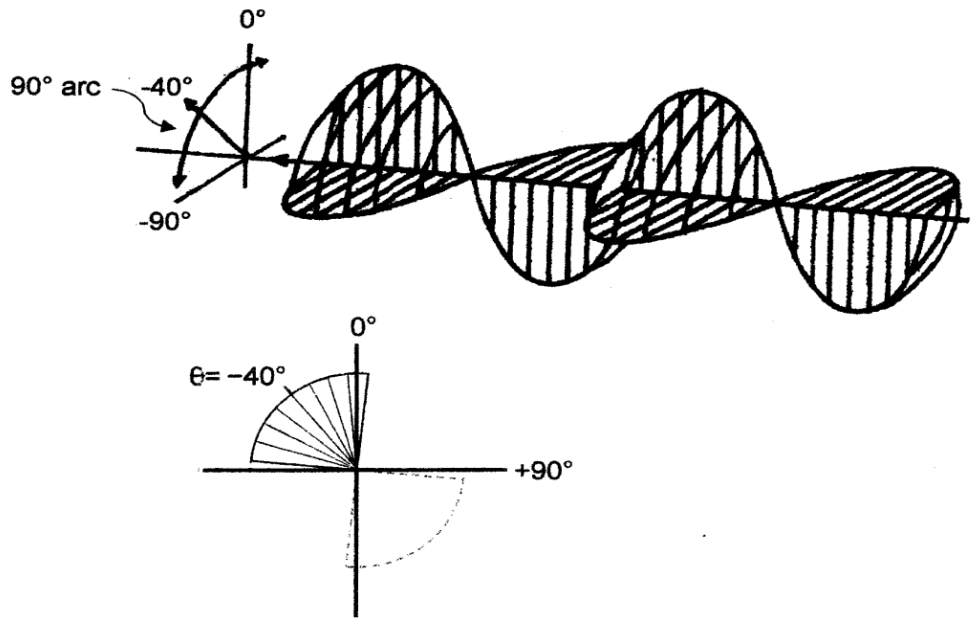

*Fig. 1. A 3-dimensional depiction of a two-wavelength segment of an objectively real linearly polarized photon wave packet and a cross-sectional view at a point along that segment showing the transverse structure of the longitudinally propagating wave packet. The depicted 90° (π/2) transverse arc is associated with an ordinary, full complement photon as well as with the generator photon of a correlated pair.*

LR specifies the interaction of a photon with a common one-channel polarizer.[6] The polarization axis of that polarizer is defined to be at 0°. A photon oriented at some random θ has a 50% chance that its wave packet arc will intersect the polarizer axis. When that intersection does occur, the wave transversely condenses onto the inclusive polarizer axis. In that process, the resident energy quantum is trapped onto the transversely condensing wave. The wave continues propagation within the polarizer as a 2-dimentional planar wave photon transversely oriented at 0°.

When the planar wave photon reaches the exit face of the one-channel polarizer a "conventional" ordinary photon with a 90° arc is emitted. That photon has a random "polarization ensemble" orientation θ'. A large number of photons exiting a linear polarizer constitute members of a polarization ensemble. The relative distribution of ensemble member orientations is given by $\cos^2\theta'$ where the polarizer axis from which it is emitted is defined is defined as 0°. This distribution is derived from first principles in [5].

Ref. [5] shows that for (polarization) correlated pairs of photons (as well as for correlated pairs of particles) there is a naturally occurring asymmetry of the wave structures of the pair members.

For photons, this asymmetry is manifested by one pair member, the "generator" photon, having a full complement $\Delta_G=\pi/2$ arc span while the "emission" photon it produces has an arc span $\Delta_E \leq \pi/2$ that ranges from 0 to $\pi/2$. The particular $\Delta_E$ value for any correlated photon pair is that of a random member of a well-defined "emission ensemble" of photons. The average arc span of the emission ensemble members is approximately 2/3 of $\pi/2$ and the emission photons are characterized as having a partial complement arc span. Additionally, the generator photon and the emission photon arcs of every pair of correlated photons have aligned bisectors and those arcs can be conveniently represented on a common-orientation reference frame as in Fig. 2a.

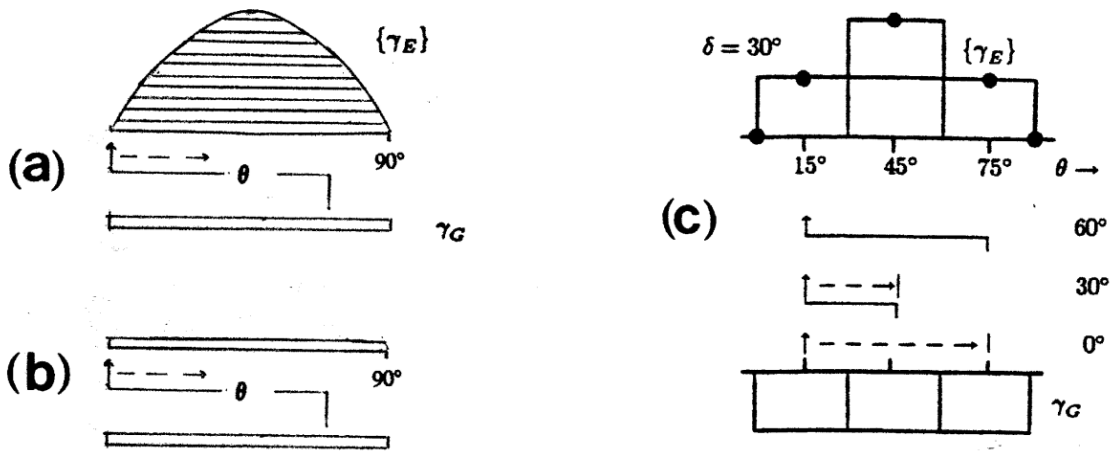

*Fig's. 2a-c. Bell measurements showing photon arc spans and relative θ rotation of analyzers preceding the detectors: (a) for a correlated generator photon $\gamma_G$ and a random member $\gamma_E$ (a row) of the analyzer emission ensemble {$\gamma_E$} defined by the sin2θ distribution envelope. (b) for a pair of identical photons extracted from a single mode. (c) for a highly simplified model of the (a) correlated pair in which the infinite member ensemble {$\gamma_E$} has only two members and arc spans for photons are partitioned in π/6 increments.*

Fig. 2a illustrates the 90° arc span of the generator member photon $\gamma_G$ of a polarization correlated photon pair. The other member is an "emission photon" $\gamma_E$ that is a random member of an analyzer (polarizer) emission ensemble {$\gamma_E$}. There is an infinite number of members of {$\gamma_E$} with a distribution of arc spans ranging from 0° to 90° defined by the rows in a sin2θ envelope. Because of the correlation relationship, a coordinate frame of reference can be set up as illustrated. The solid arrow shows the points of coincident samplings on $\gamma_G$ and {$\gamma_E$} when the respective detectors in a Bell experiment are preceded by analyzers (polarizers) relatively rotated by θ. The dashed line shows the effective scanning of that sampling process as a large number of coincident samplings are acquired. Because of the partial complement arc span of the emission member, a "correlated" pair of generator and emission photons results in the characteristic quantum mechanical joint polarizer sampling result of $\cos^2 \theta$. This result more precisely reduces to $\frac{1}{2} \cos^2 \theta$ because samplings beyond the sin2θ envelope are all null.

In contrast, as shown in Fig. 2b, a pair of ordinary full complement photons extracted from a single mode of a coherent beam are fully symmetric and identical in every sense including a common bisector orientation and a common full complement arc span. When such pairs are subjected to a Bell-experiment joint polarizer sampling the result is a linear ½(1-2θ/π) over 0 to π/2. These symmetric pairs are certainly also "correlated" but clearly distinguishable from the asymmetric generator-emission pair ½ $\cos^2 θ$ joint polarizer sampling.

Fig. 2c depicts a highly simplified model of the Fig. 2a infinite-member emission ensemble {$γ_E$} and its associated generator photon $γ_G$. In Fig. 2a the wave arcs $Δ_E$ and $Δ_G$ are properly comprised of an infinite number of angularly distributed longitudinal waves whereas in Fig. 2c the model partitions those arcs into 30° "δ wave packets" purely as a mathematical intermediary for the purposes of implementing a finite number of Bell samplings. That partitioning reduces the infinite-member emission ensemble {$γ_E$} to a mere two members, a $γ_{E1}$ comprised of a single δ wave packet and a $γ_{E2}$ comprised of three δ wave packets, the same as that of the $γ_G$. Notably, the six dots on the Fig. 2c {$γ_E$} constitute an exact fit to the sin2θ envelope of the infinite-member Fig. 2a {$γ_E$}.

The evaluation of joint transmission is totally trivial. With respect to the Fig. 2c frame, when the polarizers are not relatively rotated, θ=0° and there are 4 joint transmissions of $γ_G$ with photons in {$γ_E$}. When θ=30°, there are 3 joint transmissions. For θ=60°, there is 1 joint transmission. For θ=90° and beyond (not depicted), there are 0 joint transmissions. Normalized to 4, these results give exact agreement with ½$\cos^2 θ$. This is entirely a locally determined correlation that persists as $γ_G$ and $γ_E$ move apart. There is no non-local phenomenon of entanglement invoked in the process of achieving the correlation.

Straightforward integrations computing joint transmissions for the Bell measurements associated with the Fig. 2a continuous distribution yield the same conclusion, there is no non-local phenomenon of entanglement invoked. [5]

From the perspective of LR, asymmetric pairs as in Fig. 2a exhibit mathematical correlation levels that are directly attributable to their respective objectively real structures. Conversely, because of the explicit PI exclusion of objectively real structures, entities that are identified as being correlated by an interaction process necessarily exhibit the non-local phenomenon of entanglement upon being measured.

Finally, it is important to review the above results and their relationship to Bell's Theorem. A consequence of the asymmetry of the generator-emission pairs is that correlated emission members exhibit the property of "enhancement" when measured. In the context of photons, enhancement means that the transmission of an objectively real fully specified photon through a randomly oriented polarizer is enhanced by transmitting that photon through a second, specified polarizer. A full complement photon, such as a generator photon, has a 50% probability of being transmitted through a randomly oriented polarizer whereas for an average emission photon that probability drops to ~(2/3)50%. However, if a specified emission photon is transmitted through a specified polarizer (oriented to ensure transmission of the specified emission photon), that emission photon emerges as a full complement Δ=π/2 arc span photon with a subsequent transmission probability through a randomly oriented polarizer of 50%. As a result, transmission of the emission photon member of a correlated pair through a polarizer enhances that photon's transmission probability through a randomly oriented polarizer. A similar analysis for particles yields a corresponding outcome. [5]

Clauser and Horne deduced a general rule that locally real hidden variable theories that exhibit enhancement are not subject to Bell's Theorem. [6] However, this exclusion was not considered to be a viable loophole for locally real HVT's because the property of enhancement was conjectured to be an implausible physical property, i.e. transmitting a photon through a polarizer plausibly should not increase its transmission through a subsequent, randomly oriented polarizer. As a counterexample to this conjecture, enhancement does occur as a natural and plausible property of LR. [5] This does not constitute a conundrum with regard to Bell's Theorem and LR since we can conclude that it is the conjecture itself that is incorrect and Bell's Theorem is not applicable to LR, which is a special case of the class of HVT's that exhibit enhancement.

Because of the flawed conjecture that all physically viable HVT's had the property of non-enhancement, the reported Bell experiments in conjunction with Bell's Theorem have effectively led to the general perception that the entire class of HVT's has been disproved. A consequent corollary to this perception is that pairs of photons (or particles) that exhibit the characteristic quantum correlation are automatically inferred to be non-locally entangled. "The perception of entanglement is a consequence of the subtlety of enhancement." [5]

**SUPERPOSITION**

In this section we examine the LR representation of the photon wave function with regard to phenomenon in the probabilistic interpretation PI that are necessarily identified as non-classical superposition.

In the context of the present analysis, in which interactions with devices such as polarizers are primarily examined, the convention is followed here of using a highly compact form of the wave function that specifically treats those interactions. However, the notation used in this LR representation of wave functions is deliberately modified from that of PI in the interests of clearly distinguishing objectively real quantities from those of probabilistic quantities.

The analysis begins with a photon, an occupied electromagnetic wave mode, propagating in free space or in a non-polarized medium. The photons we consider in this section on superposition are all "ordinary" linearly polarized photons, ordinary in the sense that they have full complement 90° transverse wave arcs as opposed to the exceptional case of partial complement <90° wave arcs associated with emission correlated photons. [5] The mode is properly represented by a "pie-vector" amplitude **Φ** that is orthogonal to the propagation axis. At any point along the mode's propagation axis the mode is represented in the transverse plane by an infinite set of equal amplitude radial vectors uniformly distributed over a π/2 arc resulting in a 3-dimensional wave structure of the mode modulated along the propagation axis by the longitudinal wave function ξ. See Fig. 1. When a pie vector amplitude of the mode is incident on a polarized medium a condensation process occurs that projects the angularly distributed radial vectors onto the polarization axis (or axes) of the polarized medium. These projections result in the 3-dimensional mode condensing to a 2-dimensional mode consisting of a planar wave aligned with the medium's polarization axis in the transverse plane and modulated along the propagation axis by the longitudinal wave function ξ. Accordingly, the pie-vector **Φ** transitions to a simple vector representation **Φ**$_δ$ that is aligned to the medium's polarization axis upon entering the medium, orthogonal to the propagation axis and modulated along the propagation axis by the longitudinal wave function ξ. The δ subscript denotes that **Φ**$_δ$ is a planar wave condensed from a pie-vector **Φ** in a process analogous to a Dirac-delta function. In treating transitions between pie-vectors to vectors it is mathematically more expedient to utilize a vector form **Φ** that serves as equivalency vector substitute for the pie vector **Φ**. The requisite properties of the equivalency vector **Φ** can readily be determined by considering a **Φ**→**Φ**$_δ$ transition and imposing conservation of probability.

The pie vector form wave function representing a free space mode is

**Φ**(θ,z,t,Δ)=ξ(z,t) m **r**(θ,Δ).

The arc span parameter Δ is included in this general expression for **Φ** in free space, however, since we consider only ordinary Δ=π/2 modes in this section, Δ can be omitted. Similarly, the longitudinal wave function parameters z and t can also be omitted in the context of the present analyses within this section.

The pie vector **r**(θ) is a unit modulus, |**r**(θ)|=1, transverse representation of **Φ**(θ) comprised of an infinite set of equal amplitude unit magnitude radial vectors uniformly distributed over a π/2 arc (for an ordinary photon) with an arc bisector at θ. The longitudinal wave function is assigned a unit modulus, |ξ|=1, so that the magnitude of **Φ**(θ) is subsumed in some scalar coefficient m. The orientation θ denotes the objectively real bisector of the **Φ**(θ) arc span.

The particular example of θ=0° (defined as the vertical axis) is first examined before proceeding to a general case of θ. When an objectively real **Φ**(0°) is normally incident on a polarizer such as calcite that has a polarization axis at 0°, the cosine projections of the **r**(0°) unit magnitude radial vectors along the 0° axis integrated over ±π/4 introduces a factor of √2 which, divided by π/2, gives an average projection of almost exactly 0.9. More significantly, however, as the projective process proceeds, the resultant entity arising from the vector sum of the radial vector cosine projections is a very narrow, large amplitude peak trivially centered at 0° by symmetry in a physical process mathematically analogous to the Dirac delta function.

Physically in that process of **Φ**(0°) → **Φ**$_δ$(0°), probability is conserved consistent with the Dirac delta function applied to the squared modulus of a spatially extended amplitude condensing into a vanishingly narrow peak. The wave function of **Φ**$_δ$(0°) is represented by

$\mathbf{\Phi}_\delta(0°) = \xi\, M\, \mathbf{r}_\delta(0°)$

where δ subscripts explicitly denote the vector quantities $\mathbf{\Phi}_\delta$ and $\mathbf{r}_\delta$ as physically planar condensations in the transverse plane. M is the scalar magnitude of the $\mathbf{\Phi}_\delta$ vector amplitude.

In analyzing transitions of amplitudes into and out of polarizers it is mathematically more expeditious if all amplitudes can be expressed in vector form. To this end, the "equivalency vector" **Φ** is introduced here in mathematical replacement of the pie vector $\underline{\mathbf{\Phi}}$ which more accurately physically represents modes in free space and in non-polarized mediums.

Nevertheless, the equivalency vector **Φ** still retains the properties necessary to represent mode interactions with polarizers while properly conserving probability in those interactions.

In contrast to the pie vector,

$\underline{\mathbf{\Phi}}(0°) = \xi\, m\, \underline{\mathbf{r}}(0°)$

where its arc bisector unit vector is oriented at 0° in the transverse plane

the equivalency vector

$\mathbf{\Phi}(0°) = \xi\, M'\, \mathbf{r}(0°)$

is represented in the transverse plane by a unit radial vector **r** oriented at 0° and that vector is scaled by a magnitude M'.

In these respects the equivalency vector amplitude is structurally similar to the mode vector amplitude within a polarizer,

$\mathbf{\Phi}_\delta(0°) = \xi\, M\, \mathbf{r}_\delta(0°)$

where 0° is necessarily the orientation of the polarization axis of the polarizer in which $\mathbf{\Phi}_\delta(0°)$ is propagating. For the equivalency vector $\mathbf{\Phi}(0°)$, that δ subscript is omitted as a reminder that the equivalency vector is not physically a planar condensate but is instead a mathematical substitute for the polarizer-incident pie vector $\underline{\mathbf{\Phi}}(0°)$.

The mathematical functionality of that substitution is made possible by the probability conservation in the $\underline{\mathbf{\Phi}}(0°) \rightarrow \mathbf{\Phi}_\delta(0°)$ transition which allows, in a time reversal process, the imposition of the same magnitude onto the equivalency wave function $\mathbf{\Phi}(0°)$ as that on the planar wave function $\mathbf{\Phi}_\delta(0°)$ propagating within the polarizer, i.e. M'=M. Mathematically this imposes the probability of the pie vector $\underline{\mathbf{\Phi}}(0°)$ onto the equivalency vector $\mathbf{\Phi}(0°)$.

The above special case of θ=0° for modes incident on a polarizer can readily be extended to the general case of some objectively real θ for the pie vector arc bisector orientation. For a pie vector with an arc bisector at θ, the vector sum of the radial vectors is by symmetry a resultant vector oriented at θ.

This allows a decomposition of an equivalency vector amplitude

$$\Phi(\theta) = \Phi(0°) \cos\theta + \Phi(90°) \sin\theta$$

$$= \xi\, M \cos\theta\, \mathbf{r}(0°) + \xi\, M \sin\theta\, \mathbf{r}(90°).$$

We then consider the above decomposed vector amplitude incident on a two-channel polarizer such as calcite where the "vertical" polarization axis is at 0° and the "horizontal" polarization axis is at 90°. Specifically, the equivalency vectors $\Phi(0°) \cos\theta$ and $\Phi(90°) \sin\theta$ are incident respectively on the polarizer's 0° and 90° axes. As a result, the δ-form amplitudes

$$\Phi_\delta(0°) = \xi\, M \cos\theta\, \mathbf{r}_\delta(0°)$$

and

$$\Phi_\delta(90°) = \xi\, M \sin\theta\, \mathbf{r}_\delta(90°)$$

respectively propagate on the 0° and 90° polarization channels. The squared moduli of these two δ-form amplitudes confirms that probability is conserved in this representation and demonstrates the utility of employing the equivalency vector in place of the pie vector for the general case of a θ arc bisector.

**Summary of Notation:**

$\underline{\Phi}(\theta)$ and $\underline{\mathbf{r}}(\theta)$ are pie vectors that geometrically and mathematically correspond to modes in free space.

$\Phi(\theta)$ and $\mathbf{r}(\theta)$ are "equivalency" vectors that generally can be used as simplified mathematical vector substitutes for pie vector representations of modes in free space.

$\Phi_\delta(\theta)$ and $\mathbf{r}_\delta(\theta)$ are vector quantities that geometrically and mathematically correspond to δ-form modes in polarizers.

The outcome of energy quanta on modes incident on polarizers must also be addressed in the context of the wave functions.

The transfer of energy quanta of the mode onto a channel of a polarizer is deterministically dependent upon the intersection of the mode's incident arc intersecting that channel. That condition is determined from the equivalency vector orientation θ which is the orientation of the actual arc bisector of the 90° arc span of the "ordinary" photon wave packets we are considering here. If

$$-45° < \theta < +45°$$

the 90° transverse arc of the mode wave structure intersects the 0° vertical polarization axis and any energy quanta that occupy the mode are confined to the vertical polarization channel of the polarizer as the radial vectors of the mode projectively condense along the vertical polarization axis. Conversely if

$$+45° < \theta < +135°$$

the energy quanta are confined to the polarizer's horizontal polarization channel. For the objectively real "ordinary" photons we consider here, the arc bisector orientation of any incident photon can be fully specified as either −45°<θ<+45° or +45°<θ<+135°.

The presence of energy quanta on a mode amplitude is indicated by a "+" subscript. For example, the equivalency vector amplitude

**Φ**(θ)$_+$=ξ M **r**$_δ$(θ)$_+$= ξ **r**(θ)$_+$

is occupied and has a unit modulus since M=1.

For the case of −45°<θ<+45°, when **Φ**(θ)$_+$ is incident on a calcite polarizer, the projective condensation of that amplitude onto the vertical polarization axis along with the energy quantum residing on the wave packet arc results in an occupied δ-form vector amplitude

**Φ**$_δ$(0°)$_+$=ξ cos θ **r**$_δ$(0°)$_+$

propagating on the vertical polarization channel with a magnitude cos θ. Then, deterministically,

**Φ**$_δ$(90°)=ξ sin θ **r**$_δ$(90°)

is the vector amplitude propagating on the horizontal polarization channel with a magnitude |sin θ|. Since −45°<θ<+45°, the associated arc of the incident mode does not intersect the horizontal axis and the vector amplitude propagating on the horizontal polarization channel **Φ**$_δ$(90°) is an "empty" wave mode as indicated by the absence of a "+" subscript.

There are numerous occasions encountered in the analyses below in which an incident mode is polarized about some specified polarization axis where that axis has an orientation θ as distinct from specifying that the objectively real incident mode itself has some specified orientation. That distinction is addressed here by adopting the convention that θ$_α$ denotes the orientation of a mode that has the orientation of a random member α of a polarization ensemble centered about θ. The subscript is given by a lower-case Greek letter other than δ to avoid confusion with δ-form modes.

For example, the notation 0°$_α$ specifies that the amplitude is transversely represented by the orientation θ (bisector angle) of a random member α of the 0°-polarization ensemble. θ is in the range from -45° to +45° with a frequency distribution determined by the ensemble's cosine squared curvilinear envelope. This convention can be instructively visualized using the finite 16-member ensemble depicted in Fig. 3 where α is some random value from 1→16 and θ is the bisector angle of that α member.

An expression for the bisector orientation of a given α member can readily be computed. For the finite 16-member vertical polarization-ensemble in radians

0$_α$=cos$^{-1}$[(α-1)/15]$^{1/2}$-π/4.

The presence of an energy quantum on a discrete "photon" wave packet or of (a large number of) energy quanta on a mode is denoted by an appended subscript "+".

It is important to emphasize that any successive input mode, similarly represented by an amplitude $\underline{\Phi}_1(0°_\alpha)$, has a realized random member α orientation that is uncorrelated to the realized random member α orientation of any of the temporally preceding modes. This property is noted here because in the course of this analysis, there arise circumstances in which two or more simultaneously present amplitudes having random orientations interact with each other. For these simultaneously present amplitudes, the convention applied here is that the amplitudes have mutually non-correlated random orientations when their Greek letter indices are respectively different and have mutually correlated random orientations when those indices are respectively the same.

If incident modes are "vertically" polarized, in LR that implies that the orientation θ of the mode is statistically that of a random member of a vertical polarization ensemble as depicted in Fig. 3 using a finite-member ensemble for instructional purposes.

If the source directs these linearly polarized modes at a two-channel polarizer where the polarization axis of the ensemble is in alignment with one of the polarization axes of the polarizer, e.g. its vertical axis, then an individual source mode has an orientation $0°_\alpha$ where α is some random number 1 to 16. The present axial alignment has important consequences. From Fig. 3, the arcs of all of the ensemble member modes intersect the polarizer's vertical polarization axis.

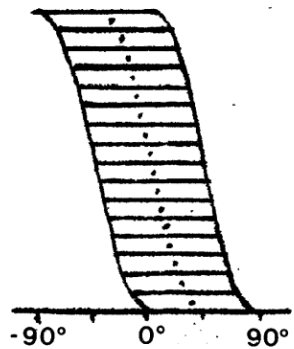

*Fig. 3. A graphical representation of a 0°-polarized ensemble of modes, showing here for instructional purposes a finite 16-member ensemble. Each member (row) represents an ordinary mode with a transverse arc span of 90°. The arc bisectors of each member, denoted by arc-centered dots, are in a statistical cosine squared distribution about 0°. The notation $0°_\alpha$ specifies a mode that has the bisector orientation of a random member α of a 0°-polarized ensemble. For the illustrated ensemble, the members are designated by α=1→16.*

As a result, for any particular mode of those 16 the energy quanta on the mode transitions onto the vertical channel of the polarizer along with the cosine projection of the incident mode amplitude. That incident amplitude from the source is

$\Phi(0°_\alpha)_+ = \xi\, r(0°_\alpha)_+$

which is normalized to a unit modulus since the magnitude M=1. The amplitude in the polarizer's vertical channel is

$\Phi_\delta(0°)_+ = \xi \cos(0°_\alpha)\, r_\delta(0°)_+$

that retains the energy quanta that had resided on the incident $\Phi(0°_\alpha)_+$ but has a reduced wave amplitude as a result of the cosine projection of the α mode onto the polarizer's vertical axis. Concurrently the amplitude on the polarizer's horizontal channel

$\Phi_\delta(90°) = \xi \sin(0°_\alpha)\, r_\delta(90°)$

has none of the energy quanta that had resided on the incident $\Phi(0°_\alpha)_+$. That empty wave amplitude has a magnitude $|\sin(0°_\alpha)|$ as a result of the sine projection of the α mode onto the polarizer's horizontal axis.

**Quantum Loop Calculations**

Quantum loops present quintessential examples demonstrating the phenomenon of non-local superposition from the perspective of "the Probabilistic Interpretation" of the underlying quantum mechanical formalism, PI. That phenomenon is often cited by PI as refutation of local realism.

The electromagnetic wave functions for two different loops are analyzed here from the perspective of LR. These loops are a pair of contiguous oppositely oriented calcite crystals as illustrated in Fig. 4 and a polarizing beam splitter Mach-Zehnder Loop, PBS M-Z Loop, as illustrated in Fig. 5. From the perspective of the Probabilistic Interpretation of quantum mechanics, PI, the quantum state of a mode to either loop input is exactly duplicated at the loop output in a process then identified as "unitary."

Mode transit of a calcite loop

The loop analyses begin with the Fig. 4 calcite loop, a pair of contiguous, oppositely oriented calcite polarizers. In this and in subsequent analyses the amplitudes are subscripted with the number of the path segment on which they reside since different path segments may be associated with consequential transitions of the wave function states.

The (unit modulus) amplitude incident on path 1 is

$\Phi_1(\theta)_+ = \xi\, r(\theta)_+$

where −45°<θ<+45° without loss of generality since analysis with the orthogonal condition, +45°<θ<+135°, is straightforward.

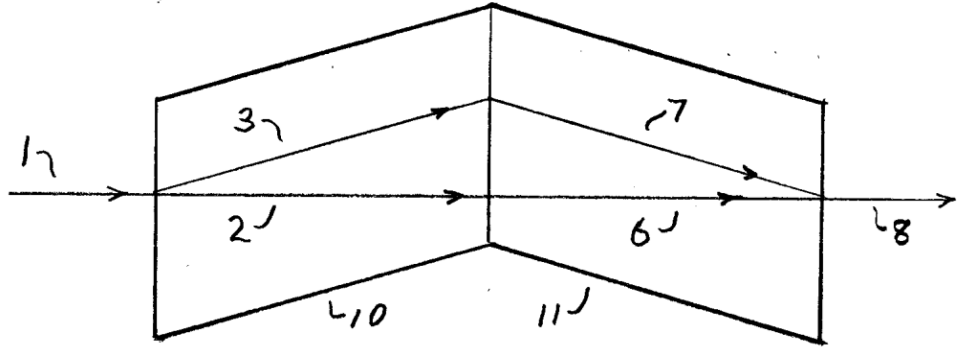

*Fig. 4. Diagram of a calcite loop.*

The orientation θ of $\Phi_1(\theta)$ may be some random value in a uniform distribution from −45° to +45°. Alternatively, $\Phi_1(\theta)$ may be identified as derived from a 0°-linearly polarized source. In that case θ is still in the range from −45° to +45° but is then derived from a statistical cosine squared distribution and the incident amplitude's orientation could be more precisely characterized as some $0°_\mu$ rather than θ. However, in the following analyses below the original distribution associated with the orientation θ of the incident $\Phi_1$ is not generally of consequence and it is more appropriate to simply identify the orientation as some objectively real θ within a range such as −45° to +45°.

Continuing onto path 2, the amplitude

$\Phi_{\delta 2}(0°)_+ = \xi \cos\theta \, \mathbf{r}_\delta(0°)_+$

is a δ-form planar wave mode because it is propagating inside a polarized medium. $\Phi_{\delta 2}$ retains the energy quanta that had been on $\Phi_1(\theta)_+$ because the transverse arc for that θ bisector orientation intersected the 0° polarization axis of calcite 10.

For path 6 the amplitude

$\Phi_{\delta 6}(0°)_+ = \Phi_{\delta 2}(0°)_+$

Resulting in

$\Phi_{\delta 6}(0°)_+ = \xi \cos\theta \, \mathbf{r}_\delta(0°)_+$

since calcites 10 and 11 are contiguous and there is no amplitude transition in the passage from the vertical 0° polarization channel on calcite 10 to that on calcite 11.

The path 3 amplitude

$\Phi_{\delta 3}(90°) = \xi \sin\theta\, \mathbf{r}_\delta(90°)$

is the complementary horizontal axis projection of the incident amplitude. $\Phi_{\delta 3}$ is an empty wave mode because of the orientation constraint $-45° < \theta < +45°$ on $\theta$. In analogy to the $\Phi_{\delta 2}(0°)_+$, $\Phi_{\delta 6}(0°)_+$ equivalence the amplitude on path 7 is

$\Phi_{\delta 7}(90°) = \xi \sin\theta\, \mathbf{r}_\delta(90°)$.

With in-phase conditions present for the two orthogonal amplitudes converging to the output face of calcite 11, the vector sum of

$\Phi_{\delta 6}(0°)_+ + \Phi_{\delta 7}(90°) = \xi\, [\cos\theta\, \mathbf{r}_\delta(0°)_+ + \sin\theta\, \mathbf{r}_\delta(90°)]$

$\qquad\qquad = \xi\, \mathbf{r}(\theta)_+$

$\qquad\qquad = \Phi_8(\theta)_+$

$\qquad\qquad = \Phi_1(\theta)_+$.

This conjuncture at the calcite 11 exit face results in a loop output mode, second line above, identical to the input mode consistent with the unitary property of the calcite loop required by the underlying quantum formalism. However, this LR representation does not impose the non-local superposition necessitated when the loop transit is represented by PI.

Mode transit of a M-Z PBS loop

Fig. 5 illustrates a Mach-Zehnder polarizing beam splitter loop. Consistent with the underlying quantum formalism, transit of this loop is unitary as is transit of a calcite loop. The LR analysis of mode transit of the M-Z PBS loop is more involved than that of the calcite loop but is nevertheless straightforward. In common with the LR transit analysis of the calcite loop, the LR transit analysis of the M-Z PBS loop does not impose the non-local superposition necessitated when the loop transit is represented by PI. From an LR perspective, the M-Z PBS loop configuration is of particular interest because of the opportunities presented by the spatial separation of the two constituent polarizers.

In the LR analysis of the M-Z PBS loop, the mode's arc bisector orientation $\theta$ is constrained to $-45° < \theta < +45°$ on the incident mode in the interests of providing an instructive example. Then

$\Phi_1(\theta)_+ = \xi\, \mathbf{r}(\theta)_+$.

Fig. 6 is diagrammatic representation of the delta-form projections of $\Phi_1(\theta)_+$ inside the PBS 25 dielectric layers, 25A. The reflected amplitude is

$\Phi_{\delta 2}(0°)_+ = \xi \cos(\theta)\, \mathbf{r}_\delta(0°)_+$

and the transmitted amplitude is

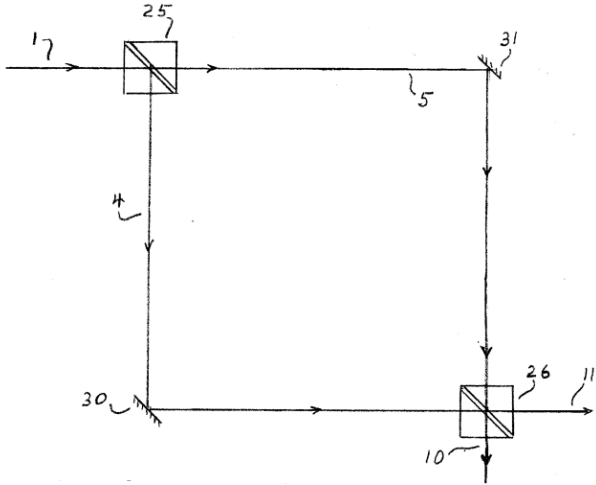

Fig. 5. A Mach-Zehnder polarizing beam splitter loop.

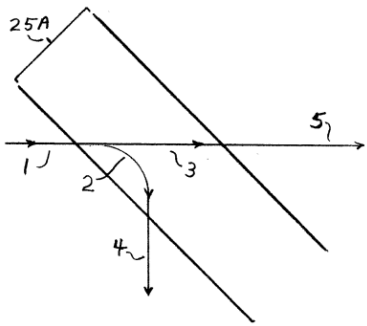

Fig. 6. A diagrammatic representation of an incident non-planar mode on path 1 projectively condensing in the PBS dielectric layer 25A to a vertically oriented delta-form planar wave on path 2 and to a horizontally oriented delta-form planar wave on path 3. Those planar waves emerge from 25A to form non-planar modes that that have the orientations of random members of vertical and horizontal polarization ensembles, respectively.

$\Phi_{\delta 3}(90°) = \xi \sin(\theta) \, r_\delta(90°)$

since the current example has -45°<θ<+45° for the incident $\Phi_1(\theta)_+$.

In a statistical emission process these amplitudes respectively produce

$\Phi_4(0°_\alpha)_+ = \xi \cos(\theta) \, r(0°_\alpha)_+$

and

$\Phi_5(0°_\alpha + 90°) = \Phi_5(90°_\alpha)$

$\qquad = \xi \sin(\theta) \, r(90°_\alpha).$

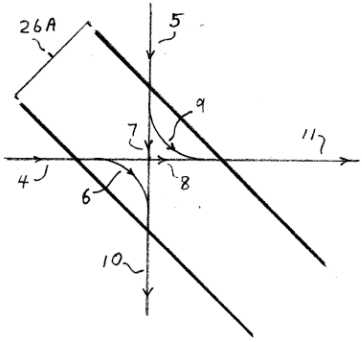

*Fig. 7. Diagrammatic representation of waves on paths 4 and 5 entering the dielectric layer 26A of PBS 26 as interacting planar delta-form waves that exit 26A as non-planar waves on paths 10 and 11. Notably, the orientations of those non-planar waves are deterministically set by the conjunction of the orthogonal planar waves from which they are formed.*

This statistical emission process occurs at the output of a polarizer when an exit wave is unaccompanied by an orthogonal exit wave.

After reflections at mirrors 30 and 31 the respective projections of $\Phi_4(0°_\alpha)_+$ and $\Phi_5(90°_\alpha)$ inside the PBS 26 dielectric layers 26A are the δ-forms

$\Phi_{\delta 6}(0°)_+ = \xi \cos(\theta) \cos(0°_\alpha) r_\delta(0°)_+$ ,

$\Phi_{\delta 7}(90°) = \xi \sin(\theta) \sin(90°_\alpha) r_\delta(90°)$

$\qquad = \xi \sin(\theta) \cos(0°_\alpha) r_\delta(90°)$,

$\Phi_{\delta 8}(0°) = \xi \cos(\theta) \sin(0°_\alpha) r_\delta(0°)$,

and

$\Phi_{\delta 9}(90°) = \xi \sin(\theta) \sin(0°_\alpha) r_\delta(90°)$.

The δ-form projections $\Phi_{\delta 6}(0°)_+$ and $\Phi_{\delta 7}(90°)$ sum vectorally to give a resultant on path 10, Fig.'s 5 and 7,

$\Phi_{\delta 6}(0°)_+ + \Phi_{\delta 7}(90°) = \xi \cos(\theta) \cos(0°_\alpha) r_\delta(0°)_+ + \xi \sin(\theta) \cos(0°_\alpha) r_\delta(90°)$

$\qquad = \xi \cos(0°_\alpha) [\cos(\theta) r_\delta(0°)_+ + \sin(\theta) r_\delta(90°)]$

$\qquad = \xi \cos(0°_\alpha) r(\theta)_+$

$\qquad = \Phi_{10}(\theta)_+$.

The line 2→3 transition of the bracketed [ ] factor to $r(\theta)_+$ represents a non-random emission process in which orthogonal, mutually accompanying waves $\Phi_{\delta 6}(0°)_+$ and $\Phi_{\delta 7}(90°)$ deterministically combine to a definite orientation.

Similarly, the δ-form projections $\Phi_{\delta 8}(0°)$ and $\Phi_{\delta 9}(90°)$ sum vectorally to give a resultant on path 11

$\Phi_{\delta 8}(0°)+\Phi_{\delta 9}(90°) = \xi \cos(\theta)\sin(0°_\alpha) r_\delta(0°) + \xi \sin(\theta)\sin(0°_\alpha) r_\delta(90°)$.

$\qquad = \xi \sin(0°_\alpha)[\cos(\theta) r_\delta(0°) + \sin(\theta) r_\delta(90°)]$

.  $\qquad = \xi \sin(0°_\alpha) r(\theta)$

$\qquad = \Phi_{11}(\theta)$.

The LR analysis of the PBS loop demonstrates that the $\Phi_{10}$ and $\Phi_{11}$ orientations are identical to that of $\Phi_1$ when $-45° < \theta < +45°$. Moreover, $|\Phi_{10}|^2 + |\Phi_{11}|^2$ shows that the output wave intensity is conserved relative to the input wave intensity $|\Phi_1|^2$. $\Phi_{10}(\theta)_+$ is occupied and $\Phi_{11}(\theta)$ is empty.

The corresponding analysis when $+45° < \theta < +135°$ is straightforward. In this case the $\Phi_1(\theta)_+$ wave packet arc intersects the horizontal polarization axis of PBS 25 resulting projective condensation of the wave packet onto that horizontal polarization axis along with the resident energy quantum. The horizontal polarization axis of a PBS is associated with transmission which relates to the δ-form path 3 in the Fig. 6 depiction of the PBS dielectric. Accordingly, $\Phi_{\delta 3}(90°)$ is occupied,

$\Phi_{\delta 3}(90°)_+ = \xi \sin(\theta) r_\delta(90°)_+$.

The accompanying $\Phi_{\delta 2}(0°)$ on δ-form path 2 is empty,

$\Phi_{\delta 2}(0°) = \xi \cos(\theta) r_\delta(0°)$.

When $\Phi_{\delta 3}(90°)_+$ exits the dielectric 25A shown in Fig. 6, we have again the same form of the wave amplitude on path 5, $\Phi_5(90°_\alpha)_+$, but occupied since the transmitted $\Phi_{\delta 3}(90°)_+$ is itself occupied.

When $\Phi_5(90°_\alpha)_+$ reaches dielectric 26A shown in Fig. 7, its $90°_\alpha$ horizontal orientation satisfies $+45° < 90°_\alpha < +135°$ and the $\Phi_5(90°_\alpha)_+$ wave arc projectively condenses onto the horizontal polarization axis of dielectric 26A carrying with it the energy quantum onto the δ-form horizontal planar wave amplitude

$\Phi_{\delta 7}(90°)_+ = \xi \sin(\theta) \cos(0°_\alpha) r_\delta(90°)_+$

where it joins with the now empty $\Phi_{\delta 6}(0°)$. Accordingly, we have the same form of the amplitude sum for $\Phi_{\delta 6}$ and $\Phi_{\delta 7}$ that we had before with the only difference being that the energy quantum is provided to the resultant $\Phi_{10}$ from path 7 instead of from path 6 giving

$\Phi_{\delta 6}(0°) + \Phi_{\delta 7}(90°)_+ = \xi \cos(\theta)\cos(0°_\alpha) r_\delta(0°) + \xi \sin(\theta)\cos(0°_\alpha) r_\delta(90°)_+$

$\qquad = \xi \cos(0°_\alpha)[\cos(\theta) r_\delta(0°)_+ + \sin(\theta) r_\delta(90°)]$

$\qquad = \xi \cos(0°_\alpha) r(\theta)_+$

$\qquad = \Phi_{10}(\theta)_+$.

Then, in summary for arbitrary θ, the M-Z PBS loop has an input incident amplitude

$\Phi_1(\theta)_+ = \xi\, r(\theta)_+$

and two outputs, the occupied

$\Phi_{10}(\theta)_+ = \xi \cos(0°_\alpha)\, r(\theta)_+$

and the empty

$\Phi_{11}(\theta) = \xi \sin(0°_\alpha)\, r(\theta)$.

These results for the LR representation show that the unit input wave intensity on $\Phi_1(\theta)_+$ is split into a $\cos^2(0°_\alpha)$ fraction on the occupied $\Phi_{10}(\theta)_+$ and a $\sin^2(0°_\alpha)$ fraction on the empty $\Phi_{11}(\theta)$. The proportionality of that split for any particular input photon is determined in a random process by the value of $0°_\alpha$ that occurs as the waves emerge from the dielectric layer (25A) of the first PBS (25). That $0°_\alpha$ value is identified as the bisector orientation of a random member of a polarization ensemble.

The fraction of the wave intensity emergent on the empty $\Phi_{11}(\theta)$ confirms that wave intensity (and probability) is conserved in LR. In contrast, PI effectively requires the non-local "collapse" of wave intensity going to path 11 and an instantaneous non-local materialization of that collapsed wave intensity going to path 10.

The transit of the M-Z PBS loop as a unitary process is understood to relate to the "observable" energy quantum but not to the wave on which it resides. This distinction is made manifest in the two output M-Z PBS loop transit but is not apparent in the simpler single output calcite loop transit.

Notably, the random process associated with the transit of the first PBS has no bearing on the path taken by the energy quantum. That path is deterministically set to be path 10 via path 4 when the objectively real input photon satisfies -45°<θ<+45° and path 10 via path 5 when the objectively real input photon satisfies +45°<θ<+135°.

Most significantly, with respect to the present section, the occupied output wave $\Phi_{10}(\theta)_+$ on path 10 has the identical θ orientation of the input wave $\Phi_1(\theta)_+$ on path 1 in a process lacking superposition.

**CONCLUSIONS**

The debate over the fundamental basis of the underlying quantum mechanical formalism, whether it is objectively real or probabilistic, continued for many years with interest at times waxing and waning following the development of that formalism in the late 1920's. For much of that time the consensus held that the debate was substantially philosophical in nature and no physical consequences would emerge from a resolution of that debate if indeed the debate could even be resolved.

This consensus outlook dissipated in the latter part of the past century with Bell's Theorem [2] followed by the results of Bell experiments [3,4] in which the probabilistic representation was strongly favored. From the certain perception that the probabilistic representation was correct, computational methods such as Shor's algorithm [1] demonstrated that extremely important physical consequences could in principle be achieved by utilizing the probabilistic phenomenon of entanglement and superposition to potentially enable the construction of a quantum computer that would exhibit a significant advantage over classical computers manifested as computational speedup.

The perception that the probabilistic interpretation is correct very much continues to persist. As a consequence, impediments encountered in achieving speedup are interpreted as technical difficulties in maintaining quantum states and not as fundamental problems with the probabilistic interpretation. This interpretation is questioned here with a locally real representation that is in agreement with performed experiments, is differentially testable against the probabilistic representation and for which entanglement and superposition are absent.